\documentclass[aps,prb,twocolumn,showpacs]{revtex4}
\usepackage{graphicx}

\begin{document}

\title{Grand-canonical variational approach for the $t-J$ model}

\author{Chung-Pin Chou$^{1}$, Fan Yang$^{2,*}$, and Ting-Kuo Lee$^{1}$}
\affiliation{$^{1}$Institute of
Physics, Academia Sinica, Nankang, Taipei 11529, Taiwan}
\affiliation{$^{2}$Department of Physics, Beijing Institute of
Technology, Beijing 100081, P. R. China}

\begin{abstract}
Gutzwiller-projected BCS wave function or the
resonating-valence-bond (RVB) state in the 2D extended $t-J$ model
is investigated by using the variational Monte Carlo technique. We
show that the results of ground-state energy and excitation spectra
calculated in the grand-canonical scheme allowing particle number to
fluctuate are essentially the same as previous results obtained by
fixing the number of particle in the canonical scheme if the grand
thermodynamic potential is used for minimization. To account for the
effect of Gutzwiller projection, a fugacity factor proposed by
Laughlin and Anderson few years ago has to be inserted into the
coherence factor of the BCS state. Chemical potential, particle
number fluctuation, and phase fluctuation of the RVB state,
difficult or even impossible to be calculated in the canonical
ensemble, have been directly measured in the grand-canonical
picture. We find that except for $La-214$ materials, the doping
dependence of chemical potential is consistent with experimental
findings on several cuprates. Similar to what has been reported by
scanning tunneling spectroscopy experiments, the tunneling asymmetry
becomes much stronger as doping decreases. We found a very large
enhancement of phase fluctuation in the underdoped regime.
\end{abstract}

\maketitle

\section{Introduction}

Since the pioneering work done by Anderson \cite{AndersonSci87}, the
superconducting (SC) state of high-$T_{c}$ cuprates has been
successfully described by the so-called $d$-wave
resonating-valence-bond ($d$-RVB) wave function with a fixed number
of particles,
\begin{eqnarray}
|\Psi_{d-RVB}^{N_{e}}\rangle=\hat{P}_{N_{e}}\hat{P}_{G}|\Psi_{d-BCS}\rangle,
\label{e:equ1}
\end{eqnarray}
where the Gutzwiller projection operator $\hat{P}_{G}$ restricts
Hilbert space without doubly occupancy at each site.
$\hat{P}_{N_{e}}$ is the projection operator onto the subspace with
$N_{e}$ electrons. According to Eq.(\ref{e:equ1}), the SC wave
function with variable particle numbers seems to be naturally
obtained by taking away $\hat{P}_{N_{e}}$. However, recently
Anderson has emphasized that a fugacity factor should be inserted in
front of the coherence factor $u_{k}$ in Eq.(\ref{e:equ1}) besides
removing $\hat{P}_{N_{e}}$ \cite{AndersonJPCS06,AndersonLTP06}. This
is the same idea as what Laughlin proposed for the Gossamer
superconductivity \cite{LaughlinPM06}. By using Gutzwiller
approximation (GA) Edegger \textit{et al.} \cite{EdeggerPRB05} have
also discussed the necessity of introducing the fugacity factor in
the grand canonical wave function. But there is no exact numerical
result to verify it. Whether GA provides an accurate fugacity factor
is still an open question.

Although Eq.(\ref{e:equ1}) has been used to explain several
important features of high-$T_{c}$ cuprates like the ground-state
phase diagram
\cite{AndersonJPCM04,YokoyamaJPSJ87,GrosPRB88,EdeggerAIP07,ParamekantiPRB04,ShihPRL04,KYYangPRB06,YokoyamaJPSJ96},
interplay between antiferromagnetism and superconductivity
\cite{ShihPRB04,ShihLTP05,HimedaPRB99}, existence of stripe states
\cite{CPChouPRB08,CPChouPRB10,HimedaPRL02}, and anomalous spectral
weights of low-lying excited states
\cite{RanderiaPRB04,Yunoki,NavePRB06,CPChouPRB06,HYYangJPCM07},
\textit{etc}., only few numerical works on the grand-canonical
$d$-RVB state have been reported so far. As far as we know, Yokoyama
and Shiba is the first to have carried out the variational
Monte-Carlo (VMC) calculations with non-conserving particle numbers
in strongly correlated Hubbard model almost two decades ago
\cite{YokoyamaJPSJ88}. By using particle-hole transformation, the
calculation can be efficiently performed. However, they did not
consider the fugacity factor in the calculation. They just briefly
mentioned how to introduce an additional variational parameter
$\alpha$ in front of the coherence factor $v_{k}$ to control the
average particle number. We also notice $\alpha$ will become very
large near half-filling and that causes serious numerical
difficulties. Thus, it is important to re-examine this approach with
the fugacity factor added in front of $u_{k}$ instead.

Using this grand-canonical wave function we can examine several
important physical quantities that were not able to obtain by fixing
number of particles. For instance, Anderson and Ong
\cite{AndersonJPCS06} showed the importance of the fugacity factor
by using the GA to calculate the famous asymmetric tunneling
conductance observed by scanning tunneling spectroscopy (STS)
\cite{RennerPRB95,HanaguriNat04,McElroyPRL05,FangPRL06}. Although in
our earlier studies \cite{CPChouPRB06} using Eq.(\ref{e:equ1}), we
have shown numerically the asymmetry arise from strong correlation
effect, this conclusion needs to be verified in the grand-canonical
calculation. Since the conductance involves the ground-state energy,
excitation spectra, and spectral weights, now the excitation
involves Bogoliubov quasi-particles unlike the excitation of
Eq.(\ref{e:equ1}) which has quasi-particles with a definite charge.
In addition, the SC order can now be calculated explicitly instead
of using the long-range pair-pair correlation function to get an
estimate. The particle number fluctuation or the phase fluctuation
in the strongly correlated SC state can be calculated directly also.
Furthermore, the doping dependence of chemical potential which was
reported by experiments \cite{HashimotoPRB08}, can be determined
directly in this grand-canonical scheme.

Below we shall first focus on computing several physical quantities
using the $d$-RVB state with fluctuating particle numbers, such as
the nearest-neighbor SC order parameter
$\Delta_{d}(\equiv\frac{1}{N}\sum_{i}\langle\hat{c}^{\dagger}_{i\uparrow}\hat{c}^{\dagger}_{i+\hat{x}\downarrow}\rangle)$,
particle number fluctuation $\Delta N_{e}(\equiv\sqrt{\langle
N_{e}^{2}\rangle-\langle N_{e}\rangle^{2}})$, chemical potential
$\mu_{g}$, and spectral weight $Z_{k\sigma}^{\pm}$, \textit{etc}.
First of all, the validity of $d$-RVB wave function including a
fugacity factor in the grand-canonical picture is numerically
confirmed, as we find the equivalence for the ground- and
excited-state features between the canonical and the grand-canonical
ensemble by optimizing the grand thermodynamic potential $F$ instead
of internal energy $E$. Next, the doping trends of chemical
potential and tunneling conductance are evaluated, which is in
agreement with the experimental observations. Both SC order
parameter and particle number fluctuation show the similar doping
dependence with a dome-like shape observed in the SC phase of
high-$T_{c}$ phase diagram. Finally, we shall examine the effect of
strong correlation on the phase field of the SC order parameter. We
find that in the underdoped region, the strong correlation effect
greatly enhances phase fluctuation in contrast with the prediction
of BCS theory.

\section{Variational Monte Carlo Method}

The Hamiltonian for the extended $t-J$ model on a two-dimensional
square lattice is given by
\begin{eqnarray}
\hat{H}&=&-\sum_{i,j,\sigma}t_{ij}\left(\tilde{c}_{i\sigma}^{\dag}\tilde{c}_{j\sigma}+h.c.\right)\nonumber\\
&+&J\sum_{\langle
i,j\rangle}\left(\mathbf{S}_{i}\cdot\mathbf{S}_{j}-\frac{1}{4}\hat{n}_{i}\hat{n}_{j}\right),
\label{e:equ2}
\end{eqnarray}
where the hopping terms $t_{ij}=t$, $t'$, and $t''$ for sites $i$
and $j$ being the nearest-, the second-nearest, and the
third-nearest-neighbors, respectively. Other notations are standard.
The SC wave function is of the form,
\begin{equation}
|\Psi_{d-RVB}\rangle=\hat{P}_{G}\prod_{\mathbf{k}}\left(\tilde{
u}_{\mathbf{k}}+\tilde{v}_{\mathbf{k}}\hat c^{\dagger}_{\mathbf{k}
\uparrow}\hat c^{\dagger}_{-\mathbf{k}
\downarrow}\right)|0\rangle,\label{gtrial}
\end{equation}
where the coefficients $\tilde{u}_{\mathbf{k}}$ and
$\tilde{v}_{\mathbf{k}}$ are not necessarily the coherence factors
in the BCS wave function.

To analyze further we turn to introduce the framework of VMC
approach. The expectation value of the Hamiltonian $\hat{H}$ using
the $d$-RVB state is evaluated as \cite{GrosAP89}
\begin{eqnarray}
E&\equiv&\frac
{\langle\Psi_{d-RVB}|\hat{H}|\Psi_{d-RVB}\rangle}{\langle\Psi_{d-RVB}|\Psi_{d-RVB}\rangle}\nonumber\\
&=&\sum_{\beta,\gamma}\rho_{\beta}\frac{\langle\beta|\hat{H}|\gamma
\rangle\langle\gamma|\Psi_{d-RVB}\rangle}{\langle\beta|\Psi_{d-RVB}\rangle}.
\label{average}
\end{eqnarray}
Here $\rho_{\beta}$ represents the Monte-Carlo sampling weight for the configuration $\beta$ defined as
\begin{equation}
\rho_{\beta}\propto\left|\langle\beta|\Psi_{d-RVB}\rangle\right|^{2}=\left|det\left(\hat{A}_{\beta}\right)\right|^{2}.
\label{weight}
\end{equation}
In the canonical ensemble, the matrix elements
$\hat{A}_{\beta}(i,j)$ is given by
\begin{equation}
\hat{A}_{\beta}(i,j)=\sum_{\mathbf{k}}\frac{\tilde{v}_{\mathbf{k}}}{\tilde{u}_{\mathbf{k}}}e^{i\mathbf{k}\cdot\left(\mathbf{R}_{i\uparrow}-\mathbf{R}_{j\downarrow}\right)},\label{matrixc}
\end{equation}
where $\mathbf{R}_{i\uparrow}$ and $\mathbf{R}_{j\downarrow}$ is the
position vector of the i-th up-spin electron and the j-th down-spin
electron in the configuration $\beta$, respectively. In
Eq.(\ref{gtrial}), the particle number fluctuation results in the
unknown matrix size bringing technical difficulty to VMC
calculations. To resolve this problem, a spin-dependent
particle-hole transformation has to be introduced.

Following Yokoyama and Shiba \cite{YokoyamaJPSJ88} we introduce the
partial particle-hole transformation to change the original
representation (c) to a new representation (df) expressed as
\begin{eqnarray}
\hat{c}_{i\uparrow}&\rightarrow&\hat{f}_{i},\nonumber\\
\hat{c}_{i\downarrow}&\rightarrow&\hat{d}_{i}^{\dagger},\label{phtr}
\end{eqnarray}
where only the down-spin electrons are transformed. Here we
introduce two different particles, $d$ and $f$, instead of down- and
up-spin electrons. Both operators $\hat{f}_{i}$ and $\hat{d}_{i}$
also satisfy the anti-communication relation. Thus, three possible
Fock states at each site can be transformed in the following way,
\begin{eqnarray}
|0\rangle_{(c)}&\rightarrow&|d\rangle_{(df)}\nonumber\\
|\downarrow\rangle_{(c)}&\rightarrow&|0\rangle_{(df)}\nonumber\\
|\uparrow\rangle_{(c)}&\rightarrow&|df\rangle_{(df)}.\label{phtrstate}
\end{eqnarray}
The subscripts indicate different representations. Now
Eq.(\ref{gtrial}) can be transformed into the representation (df),
\begin{eqnarray}
&&\hat{P}_{G}\prod_{\mathbf{k}}\left(\tilde{u}_{\mathbf{k}}+\tilde{v}_{\mathbf{k}}\hat{c}^{\dagger}_{\mathbf{k}\uparrow}\hat{c}^{\dagger}_{-\mathbf{k}\downarrow}\right)|0\rangle_{(c)}\nonumber\\
&\rightarrow&\hat{P}_{G}\prod_{\mathbf{k}}\left(\tilde{u}_{\mathbf{k}}\hat{d}^{\dagger}_{\mathbf{k}}+\tilde{v}_{\mathbf{k}}\hat{f}^{\dagger}_{\mathbf{k}}\right)|0\rangle_{(df)}.\label{gtriald}
\end{eqnarray}

The Gutzwiller projection operator $\hat{P}_{G}$ in the
representation (df) restricts the Hilbert space to the three
possible states shown in Eq.(\ref{phtrstate}). Eq.(\ref{gtriald})
displays a quantum state that the total number of $d$ and $f$
particles is fixed to the lattice size $N$. Thus there is no total
particle number fluctuation in the representation (df). This
conservation can be understood from Eq.(\ref{phtrstate}) as well. It
suggests that for total $S_{z}=0$ the number of empty and
doubly-occupied sites are always equal in the representation (df),
implying the total number of $d$ and $f$ particles is exactly equal
to $N$. However, the number of $d$ and $f$ particles can vary even
though the sum is fixed. This fluctuation replaces the particle
number fluctuation in the original representation (c). The coherence
factors in Eq.(\ref{gtriald}) determine the number of $d$ or $f$.
Notice that in the representation (df) the average number difference
between the particles $d$ and $f$ is equal to doping density
$\delta$.

In the canonical ensemble we usually have two Monte Carlo processes:
hopping and exchanging. In other words, an up or down spin can hop
to a hole site and two anti-parallel spins can exchange with each
other. We can generate all states by applying one or both of the
processes sequentially. To connect the Hilbert spaces with different
particle numbers in the grand-canonical scheme, however, we have to
consider a new process to create or annihilate pairs. To change from
a configuration with $N_{e}$ particles to another with $N_{e}\pm2$
particles, we randomly choose two different sites $i$ and $j$. If
the sites $i$ and $j$ have opposite spins, we destroy the pair at
these two sites. Conversely, if both sites $i$ and $j$ are empty, we
create a pair. In the representation (df) these processes also can
be easily implemented by changing two sites both with a single $d$
particle to an empty site and a site doubly occupied with $d$ and
$f$ particles or vice versa.

\section{The $d$-RVB Wave Function in the Grand-Canonical Ensemble}

According to the earlier results from GA \cite{EdeggerPRB05}, they
showed that Gutzwiller projection not only imposes a local
constraint on the electron number at each site, but globally
influences the total number of electrons. In the following, we will
provide another argument to obtain the $d$-RVB wave function with
non-conserving particle numbers. Since there exists particle number
fluctuation $\Delta N_{e}(\propto\sqrt{N})$ stemming from the SC
order, the electron number in the SC ground state will have a
distribution, $\rho\left(N_{e}\right)$. The distribution function
for the wave function without Gutzwiller projection,
$\rho_{0}(N_{e})$, has an approximate Gaussian form
$e^{-\left(N_{e}-\bar{N}_{e}^{(0)}\right)^{2}/\left(\Delta
N_{e}^{(0)}\right)^2}$, centered at the most probable number of
electrons $\bar{N}_{e}^{(0)}$ with a width $\Delta N_{e}^{(0)}$. In
the thermodynamic limit, $\bar{N}_{e}^{(0)}$ is the average number
of electrons. We define $N_{e}\equiv N\left(1-\delta\right)$ and
$\Delta N_{e}^{(0)}\equiv\sigma_{0} \sqrt N$ so that the variable
$N_{e}$ in $\rho_{0}(N_{e})$ can be changed into the doping
$\delta$,
\begin{equation}
\rho_{0}(\delta)\propto
e^{-N\left(\delta-\delta_{0}\right)^{2}/\sigma_{0}^2}.
\label{distrib}
\end{equation}
The distribution function of the Gutzwiller-projected wave function
$\rho(\delta)$, reduced by a weighting factor $W_{\delta}$, deviates
from $\rho_{0}(\delta)$ due to the constraint of no-doubly occupancy
\cite{EdeggerPRB05}. This weighting factor can be estimated roughly
by counting the number of configurations in the phase space. Before
Gutzwiller projection, there are $C^{N}_{N\left(1-\delta\right)/2}$
choices for the up-spin or down-spin electrons to give us a total
configuration number
$N_{b}\left(=(C^{N}_{N\left(1-\delta\right)/2})^2\right)$. After the
projection, the total number is changed into $N_{a}\left(
=C^{N}_{N\left(1-\delta\right)/2}\cdot
C^{N\left(1+\delta\right)/2}_{N\left(1-\delta\right)/2}\right)$.
Thus, the weighting factor can be written as
\begin{eqnarray}
W_{\delta}=\frac{N_{a}}{N_{b}}&=&\frac{\left[\left(N(1+\delta)/2\right)!\right]^2}{N!\left(N\delta\right)!}\nonumber\\
&=&\frac{\left[\left(N(1+\delta)/2\right)!\right]^2}{\left[(N/2)!\right]^{2}\left(N\delta\right)!}\cdot
W_{0}.\label{portion}
\end{eqnarray}

Now we assume $N$ is very large. By using Stirling's approximation
$\left(N!\approx N^{N}/e^{N}\right)$, Eq.(\ref{portion}) can be
reduced to
\begin{equation}
W_{\delta}\simeq\left(\frac{1+\delta}{2\delta}\right)^{N\delta}\cdot\left(\frac{1+\delta}{2}\right)^{N}.\label{portion3}
\end{equation}
Eqs.(\ref{distrib}) and (\ref{portion3}) lead to the distribution
function $\rho(\delta)$
\begin{eqnarray}
\rho(\delta)&=&\rho_{0}(\delta)\cdot W_{\delta}\nonumber\\
&\propto&e^{-N\left(\delta-\delta_{0}\right)^{2}/\sigma_{0}^2}\left(\frac{1+\delta}{2\delta}\right)^{N\delta}\left(\frac{1+\delta}{2}\right)^{N}.\label{distrib2}
\end{eqnarray}
This function is very different from $\rho_{0}(\delta)$. Not only
the maximum of the distribution function $\rho(\delta)$ is shifted
from $\delta_{0}$ to the higher doping density $\bar{\delta}$, the
width also becomes considerably narrower. In the thermodynamic
limit, $\bar{\delta}$ in the distribution function is just the
average doping density $\langle\delta\rangle$. Thus, the doping
density in the grand-canonical ensemble is determined not only by
variational parameters but also the Gutzwiller projection operator.
Similar conclusions have also been discussed previously
\cite{AndersonIJMPB11,EdeggerPRB05}.

\begin{figure}[top]
\rotatebox{0}{\includegraphics[height=3.6in,width=3.4in]{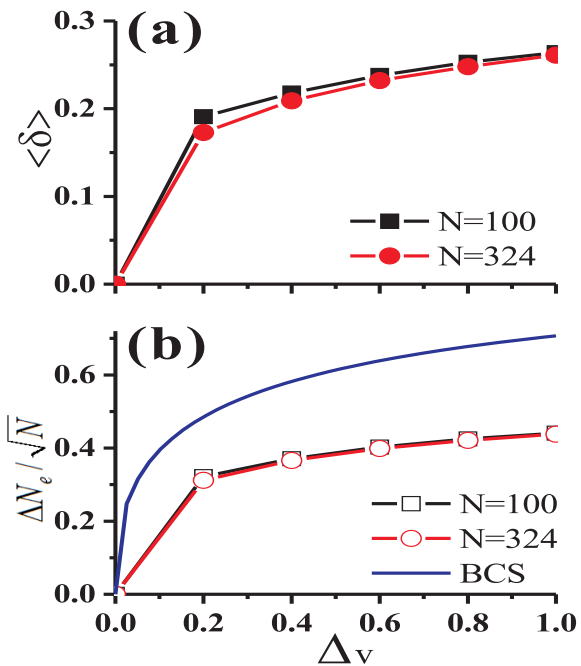}}
\caption{(Color online) (a) The average doping density and (b) its
fluctuation as a function of the variational parameter $\Delta_{v}$
for the wave function of Eq.(\ref{gtriald}) using the BCS coherence
factors. Results for a cluster of $10\times10$ and $18\times18$ are
shown in black and red, respectively. Other parameters are set to be
$\mu_{v}=t_{v}'=t_{v}''=0$. The blue line in (b) is the $d$-wave BCS
result.} \label{fig.1}
\end{figure}

We can further study Eq.(\ref{distrib2}) by carrying out the
numerical calculation using the VMC method. We consider the simple
BCS wave function by replacing $\tilde{u}_{\mathbf{k}}$ and
$\tilde{v}_{\mathbf{k}}$ in Eq.(\ref{gtriald}) with $u_{\mathbf{k}}$
and $v_{\mathbf{k}}$, respectively. $u_{\mathbf{k}}$ and
$v_{\mathbf{k}}$ are the BCS coherence factors, defined by
\begin{eqnarray}
u_{k}&=&\sqrt{\frac{1}{2}\left(1+\frac{\epsilon_{k}}{\sqrt{\epsilon_{k}^{2}+\Delta_{k}^{2}}}\right)},\nonumber\\
v_{k}&=&Sgn(\Delta_{k})\sqrt{\frac{1}{2}\left(1-\frac{\epsilon_{k}}{\sqrt{\epsilon_{k}^{2}+\Delta_{k}^{2}}}\right)},\label{uvk}
\end{eqnarray}
where
\begin{eqnarray}
\epsilon_{k}&=&-2\left(cosk_{x}+cosk_{y}\right)-4t_{v}'cosk_{x}cosk_{y}\nonumber\\
&&-2t_{v}''\left(cos2k_{x}+cos2k_{y}\right)-\mu_{v},\nonumber\\
\Delta_{k}&=&2\Delta_{v}\left(cosk_{x}-cosk_{y}\right).\label{edk}
\end{eqnarray}
Here $\Delta_{v}$, $\mu_{v}$, $t_{v}'$, and $t_{v}''$ are
variational parameters. For illustration we will consider the
half-filling wave function with the variational parameters:
$\mu_{v}=t_{v}'=t_{v}''=0$. The only parameter left is $\Delta_{v}$
which varies from $0$ to $1$. In Fig.\ref{fig.1}(a) and (b), we show
the average doping density $\langle\delta\rangle(\equiv1-N_{e}/N)$
and the particle number fluctuation $\Delta N_{e}/\sqrt{N}$ as a
function of $\Delta_{v}$, respectively. The result of the $d$-wave
BCS state without projection is also shown in Fig.\ref{fig.1}(b).
Clearly as discussed above the introduction of projection has
greatly changed the distribution function of particle number with a
larger average doping density and smaller fluctuation.

Now we can construct the $d$-RVB wave function in the
grand-canonical ensemble by taking into account this change of
distribution function. Eq.(\ref{portion3}) suggests that the
important doping dependence of the distribution function in the
presence of the Gutzwiller projection operator $\hat P_{G}$ is the
factor $g^{-\hat{N}_{h}}$, where $g\equiv\frac{2\delta}{1+\delta}$,
$\hat{N}_{h}=N-\hat{N}_{e}$, and
$\hat{N}_{e}=\sum_{i,\sigma}\hat{c}_{i\sigma}^{\dag}\hat{c}_{i\sigma}$.
The operator $g^{-\hat{N}_{h}}$ mimicking the effect of
$\hat{P}_{G}$ tends to increase the doping density away from
half-filling. To balance this effect, we have to place the operator
$\sqrt{g}^{\hat{N}_{h}}$ in front of the $d$-RVB wave function.
Thus, the $d$-RVB state in the grand-canonical scheme can be written
as
\begin{eqnarray}
|\Psi_{d-RVB}^{g}\rangle&=&\hat{P}_{G}\sqrt{g}^{\hat{N}_{h}}\prod_{\mathbf{k}}\left(u_{\mathbf{k}}+v_{\mathbf{k}}\hat{c}^{\dagger}_{\mathbf{k}\uparrow}\hat{c}^{\dagger}_{-\mathbf{k}\downarrow}\right)|0\rangle_{(c)}\nonumber\\
&\rightarrow&\hat{P}_{G}\prod_{\mathbf{k}}\left(gu_{\mathbf{k}}\hat{d}^{\dagger}_{\mathbf{k}}+v_{\mathbf{k}}\hat{f}^{\dagger}_{\mathbf{k}}\right)|0\rangle_{(df)}\nonumber\\
&\equiv&\hat{P}_{G}|\Psi_{d-BCS}^{g}\rangle. \label{RVB}
\end{eqnarray}
Now $g$ can be seen as a variational parameter like $\mu_{v}$,
$t_{v}'$, $t_{v}''$, and $\Delta_{v}$. Eq.(\ref{RVB}) with the
fugacity factor is exactly the wave function proposed by Anderson
\cite{AndersonJPCS06} and Laughlin \cite{LaughlinPM06}. Previously
the fugacity factor $g$ was estimated to be
$\frac{2\delta}{1+\delta}$ by using GA.

\begin{figure}[top]
\rotatebox{0}{\includegraphics[height=3.9in,width=3.6in]{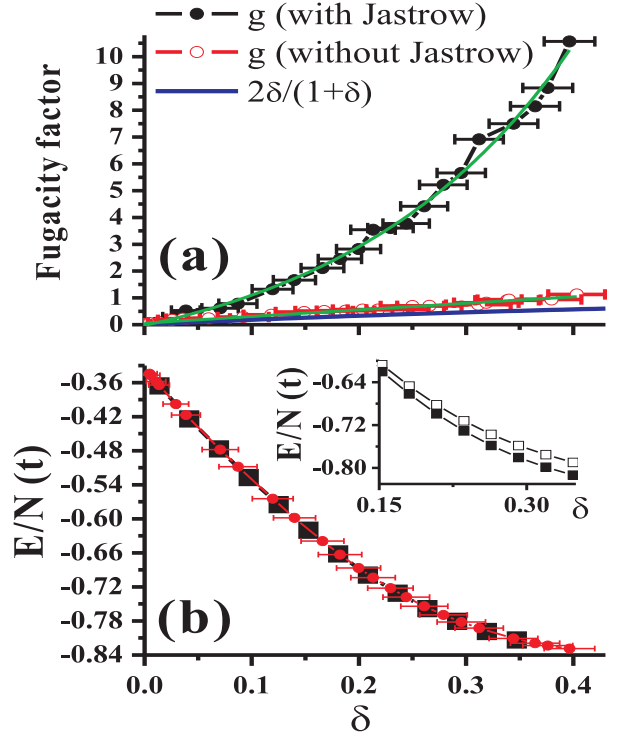}}
\caption{(Color online) The doping dependence of (a) the variational
parameter $g$ and (b) the optimized energy per site for a
$12\times12$ lattice. In (a), Solid (Empty) circles indicate the
fugacity factors $g$ with (without) Jastrow factors. The blue line
shows the renormalized Gutzwiller's factor
$\frac{2\delta}{1+\delta}$. The green lines are the guide to the
eyes. In (b), black squares (Red circles) indicate the $d$-RVB wave
function with Jastrow factors in the (grand-) canonical ensemble.
Error bars represent the average density. Inset: Solid (Empty)
squares represent the $d$-RVB wave function with (without) Jastrow
factors in the canonical ensemble. The bare parameters in the
Hamiltonian are set to be $(t',t'',J)/t=(-0.3,0.15,0.3)$.}
\label{fig.2}
\end{figure}

To obtain the ground state in the grand-canonical case, we have to
optimize the grand thermodynamic potential $F=E-\mu_{g}N_{e}$
instead of the internal energy $E$ with a fixed chemical potential
$\mu_{g}$. Here $N_{e}$ is the average number of electrons for each
$\mu_{g}$. Notice that $\mu_{g}$ is different from the variational
parameter $\mu_{v}$ in the $d$-RVB wave function. To have a lower
ground-state energy, for all the numerical results discussed below
we consider a modified $d$-RVB wave function
$\hat{P}_{J}|\Psi_{d-RVB}^{g}\rangle$ where we introduce a hole-hole
repulsive Jastrow factor $\hat{P}_{J}$
\cite{HellbergPRL91,ValentiPRL92,SorellaPRL02,CPChouPRB08}:
\begin{eqnarray}
\hat{P}_{J}=\prod_{i<j}\left[1-\left(1-r_{ij}^{\alpha}v_{\beta}^{\delta_{j,i+\beta}}\right)\hat{n}_{i}^{h}\hat{n}_{j}^{h}\right]\label{jf}
\end{eqnarray}
with
\begin{eqnarray}
r_{ij}=\sqrt{\sin^{2}\left(\frac{\pi}{L}(x_{i}-x_{j})\right)+\sin^{2}\left(\frac{\pi}{L}(y_{i}-y_{j})\right)}.\label{rij}
\end{eqnarray}
Here $\hat{n}_{i}^{h}=1-\sum_{\sigma}c_{i\sigma}^{\dag}c_{i\sigma}$.
The three parameters $v_{\beta}$ with $\beta$ to be the nearest,
second nearest, and third nearest neighbors are for short-range
hole-hole repulsion if these values are less than 1. The factor
$r_{ij}^{\alpha}$ is for long-range correlations and it is repulsive
if $\alpha$ is positive. $L$ is the linear scale of the lattice. In
Fig.\ref{fig.2}(a), we obtain the doping dependence of the fugacity
factor $g$ from the grand-thermodynamic-potential optimization.
Interestingly, since Jastrow factors will reduce the probability
that holes come closer, the fugacity factor of the modified $d$-RVB
state is greatly enhanced as increasing doping. If we do not
consider Jastrow factors, the doping dependence of $g$ is similar to
the renormalized Gutzwiller's factor $2\delta/(1+\delta)$.
Therefore, although the GA result is not exact, it is a reasonable
approximation.

\section{Quasi-particle Energy Dispersion and Spectral Weight}

In Fig.\ref{fig.2}(b), the optimized internal energy per site is
plotted as a function of the doping density for both canonical
(black squares) and grand canonical (red circles) cases. The results
are essentially indistinguishable for a $12\times12$ lattice. Thus
the ground-state phase diagram obtained in the canonical ensemble
\cite{ShihPRL04,KYYangPRB06} will be essentially the same as the
grand-canonical scheme. This result is expected as the calculation
of internal energy only involves states with the same number of
particles \cite{AndersonIJMPB11}. Additionally, the energies
obtained with the Jastrow factor are quite lower than without the
factors for the canonical case, as shown in the inset of
Fig.\ref{fig.2}(b). The fugacity factor $g$ of the modified $d$-RVB
state is very different from the case without Jastrow factors shown
in Fig.\ref{fig.2}(a).

It should be noted that we can also use the results in
Fig.\ref{fig.2}(b) obtained by Eq.(\ref{e:equ1}) to calculate the
relation between chemical potential and average number of particles
as $\mu_{g}\equiv\partial E/\partial N_{e}$. Completely consistent
results are obtained. The excellent agreement between the energies
calculated by the grand-canonical and canonical schemes is the most
important numerical result of this paper to firmly establish the
validity of inclusion of the fugacity factor $g$ in Eq.(\ref{RVB})
and our Monte Carlo algorithm. Now we can start to calculate
excitation spectra and other physical quantities in the
grand-canonical ensemble.

Not only we shall consider the energy dispersion of the excitations
but also the spectral weight below. These results could be compared
with the measured STS. The simplest way to construct a
single-particle excitation from the $d$-RVB state is to bring the
quasi-particle creation operator
$\tilde{\gamma}_{k\sigma}^{\dag}(=\frac{gu_{k}}{\sqrt{(gu_{k})^{2}+v_{k}^{2}}}c_{k,\sigma}^{\dag}-\frac{v_{k}}{\sqrt{(gu_{k})^{2}+v_{k}^{2}}}c_{-k,-\sigma})$
into play,
\begin{eqnarray}
|\Psi_{k\sigma}^{g}\rangle=\hat{P}_{G}\tilde{\gamma}_{k\sigma}^{\dag}|\Psi_{d-BCS}^{g}\rangle.\label{exc}
\end{eqnarray}
Excitation energies cannot be calculated from the internal energy
difference $E_{k}-E_{0}$ as in the canonical ensemble
\cite{CPChouPRB06}, where $E_{0}$ is the ground-state energy. Here
the chemical potential is fixed, hence we must calculate the grand
thermodynamic potential for the ground state and excited states,
$F_{0}$ and $F_{k}$, respectively,
\begin{eqnarray}
F_{0}&=&\langle\Psi_{d-RVB}^{g}|\hat{H}-\mu_{g}\hat{N}_{e}|\Psi_{d-RVB}^{g}\rangle,\nonumber\\
F_{k}&=&\langle\Psi_{k\sigma}^{g}|\hat{H}-\mu_{g}\hat{N}_{e}|\Psi_{k\sigma}^{g}\rangle.\label{free}
\end{eqnarray}
It is noticed that due to the BCS coherence factors $u_{k}$ and
$v_{k}$, the excited state $|\Psi_{k\sigma}^{g}\rangle$ will have
fewer (more) average particle numbers below (above) the Fermi level
than the ground state $|\Psi_{d-RVB}^{g}\rangle$ (not shown). Here
we have made an important assumption that these particular states
Eq.(\ref{exc}) have little overlap with other states with same
momentum and spin. This is probably valid for low-energy excitations
less than the gap energy.

\begin{figure}[top]
\rotatebox{0}{\includegraphics[height=3.5in,width=3.4in]{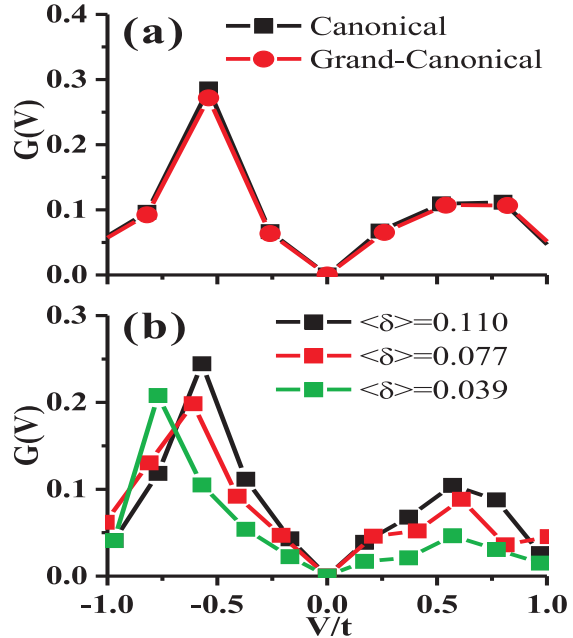}}
\caption{(Color online) $G(V)$ for the $d$-RVB state versus the bias
$V$. (a) Canonical (Grand-canonical) results for hole doping
$\delta$ (chemical potential $\mu_{g}$) fixed to $0.125$ ($1.68t$)
in $12\times12$ lattice. (b) Grand canonical results for three
chemical potentials $\mu_{g}$, $1.68t$ (black-squares), $1.9t$
(red-circles), and $2.1t$ (green-triangles). $V$ is negative
(positive) for removing (adding) one electron in $20\times20$
lattice. The bare parameters in the Hamiltonian are set to be
$(t',t'',J)/t=(-0.3,0.15,0.3)$.} \label{fig.3}
\end{figure}

The above quasi-particle states are then used to calculate the
spectral weight for adding (removing) one electron with momentum $k$
($-k$) and spin $\sigma$ ($-\sigma$) to the ground state as defined
\begin{eqnarray}
Z_{k,\sigma(-k,-\sigma)}^{+(-)}=\frac{\left|\langle\Psi_{k\sigma}^{g}|c_{k,\sigma}^{\dag}(c_{-k,-\sigma})|\Psi_{d-RVB}^{g}\rangle\right|^{2}}{\langle\Psi_{d-RVB}^{g}|\Psi_{d-RVB}^{g}\rangle\langle\Psi_{k\sigma}^{g}|\Psi_{k\sigma}^{g}\rangle}.\label{sw}
\end{eqnarray}
Similar to what we have done for the canonical case
\cite{CPChouPRB06}, we could also calculate the tunneling
conductance for the grand-canonical wave function with the bias
given by $V=F_{k}-F_{0}$. For the numerical calculations, we define
the tunneling conductance as
\begin{eqnarray}
G(V)=\frac{1}{N\Delta F}\sum_{k\in V\pm\Delta F/2}Z_{\pm
k,\pm\sigma}^{\pm},\label{gc}
\end{eqnarray}
where $\Delta F=0.28t \ (0.2t)$ for $N=144 \ (400)$ chosen is to
reduce the effect due to the finite lattice size and $\sigma$ either
up or down spin. Similar to Fig.\ref{fig.2}(b), Fig.\ref{fig.3}(a)
also shows except for high voltage the tunneling conductances are
almost identical in the canonical and the grand-canonical ensemble.
Once again, the equivalence between the canonical and the
grand-canonical ensemble for the excitation spectra within the gap
convinces us of the conclusion that the modified $d$-RVB wave
function with the fugacity factor is the precise representation for
the RVB-type state in the grand-canonical case.

\begin{figure}[top]
\rotatebox{0}{\includegraphics[height=2.in,width=3.4in]{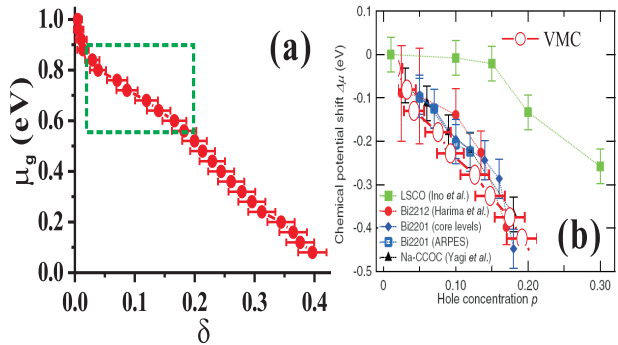}}
\caption{(Color online) (a) The doping dependence of chemical
potential $\mu_{g}$ in a $12\times12$ lattice. Red circles represent
the $d$-RVB state with Jastrow factors in the grand-canonical
ensemble. Error bars indicate $\Delta N_{e}/N$ at each average
doping. (b) The doping dependence of chemical-potential shift from
experiments in several cuprate materials \cite{HashimotoPRB08}
compared with the region shown by the green-dashed frame in (a). Red
empty circles in (b), shifted along the vertical direction for
comparison, are the data from the green-dashed frame in (a). The
bare parameters in the Hamiltonian are set to be
$(t',t'',J)/t=(-0.3,0.15,0.3)$. Here $t$ is set to be $0.4eV$.}
\label{fig.4}
\end{figure}

In addition, to numerically examine the particle-hole asymmetry of
tunneling spectra in the grand-canonical ensemble, we present the
doping dependence of $G(V)$ in Fig.\ref{fig.3}(b). The asymmetry is
clearly observed for all three doping densities. At $\delta=0.039$,
the total spectral weight for removing one electron, defined as the
sum of $Z_{k\sigma}^{-}$ over the Brillouin zone, is found to be
about three times as large as the one for adding one electron. The
gap value deduced from the width between peak positions decreases
with doping. Note that the gap size is usually overestimated in the
extended $t-J$ models. However, the weight value obtained from the
peak height enhances as increasing doping, apparently
anticorrelating with the gap size. All of these features have been
shown in the canonical ensemble \cite{CPChouPRB06} and qualitatively
consistent with the STS measurements \cite{GomesNat07,PushpSci09}.
Although the details of the tunneling spectra in
Ref.~\onlinecite{AndersonLTP06} is not completely identical to that
shown in Fig.\ref{fig.3}(b), their important characteristics are
similar.

\section{Chemical Potential}

The doping dependence of $\mu_{g}$ calculated in the grand-canonical
ensemble is plotted in Fig.\ref{fig.4}(a). It decreases
monotonically with doping as expected from Fig.\ref{fig.4}(b). Near
half-filling, $\mu_{g}$ seems to increase greatly. The
compressibility becomes extremely small as electron density
approaches half filling, this is a consequence of the opening of a
charge gap of a Mott insulator. This behavior was also observed in
the calculation of Yokoyama and Shiba \cite{YokoyamaJPSJ88}. But the
quantitative behavior is different as the fugacity factor was not
included in their grand-canonical wave function. We have also
examined the doping dependence of $\mu_{g}$ for several bare
parameters $t'/t$. The slopes for all $t'/t$ are similar except for
very large doping (not shown). Here, only the relative value of
$\mu_{g}$ is meaningful. Thus we can shift $\mu_{g}$ to compare with
experiments. Figure \ref{fig.4}(b) shows chemical-potential shift
for several cuprates observed by photoemission experiments
\cite{HashimotoPRB08}. Except for $La-214$ samples the slope of
chemical potential as a function of $\delta$ obtained by our
calculation (see the green-dashed frame in Fig.\ref{fig.4}(a)) is
almost identical to the experimental data. This result is more
quantitatively reliable in comparison with the experiments than our
previous studies in the canonical ensemble \cite{KYYangPRB06} in the
low doping region. As for the inconsistency with $La-214$ cuprates,
a possible reason could be the existence of stripe order in those
materials \cite{ZaanenAP96,InoPRL97}. It will be interesting to
investigate chemical-potential shift by using the
Gutzwiller-projected stripe wave function \cite{CPChouPRB10} in the
grand-canonical ensemble in the future.

\begin{figure}[top]
\rotatebox{0}{\includegraphics[height=2.5in,width=3.4in]{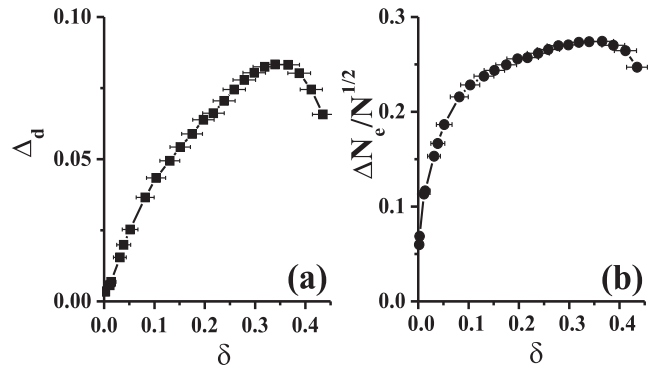}}
\caption{The doping dependence of (a) the SC order parameter
$\Delta_{d}$ and (b) the scaled particle number fluctuation $\Delta
N_{e}/\sqrt{N}$ using the $d$-RVB state in the $12\times12$ lattice
in the grand-canonical ensemble. Error bars indicate $\Delta
N_{e}/N$ at each average doping. The bare parameters in the
Hamiltonian are set to be $(t',t'',J)/t=(-0.2,0.1,0.3)$.}
\label{fig.5}
\end{figure}

\section{SC Order Parameter and Particle Number Fluctuation}

In order to investigate the SC characteristics of the $d$-RVB wave
function, we calculate the SC order parameter $\Delta_{d}$ and the
particle number fluctuation $\Delta N_{e}$ impossible to be derived
in the canonical ensemble, as shown in Fig.\ref{fig.5}(a) and (b),
respectively. The SC order parameter $\Delta_{d}$ which is
presumably proportional to the SC critical temperature $T_{c}$ is
plotted as a function of doping density in Fig.\ref{fig.5}(a). The
dome-like shape comparable to the SC phase in the cuprate phase
diagrams is consistent with earlier VMC results in the canonical
ensemble \cite{CPChouPRB06}. The density with the maximum value of
order parameter is again much larger than the optimal doping density
in cuprates. In the BCS theory the fluctuation of the particle
number, $\Delta N_{e}$, is proportional to the SC order parameter.
In Fig.\ref{fig.5}(b), $\Delta N_{e}/\sqrt{N}$ is plotted as a
function of doping density. Comparison between Fig.\ref{fig.5}(a)
and (b) shows that both $\Delta N_{e}$ and $\Delta_{d}$ calculated
by using the $d$-RVB state have the similar dome-like shape, but
there is a significant difference at underdoped regime. This
difference is clearly due to the projection operator or the Mott
physics as the particle number fluctuation is suppressed for low
doping.

\section{Phase Fluctuation}

\begin{figure}[top]
\rotatebox{0}{\includegraphics[height=2.8in,width=3.5in]{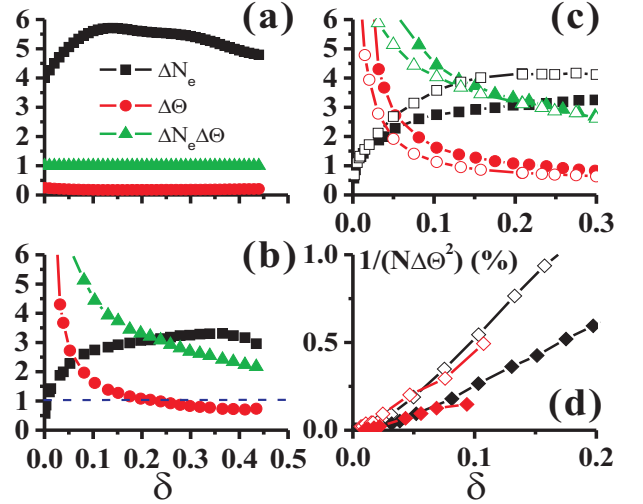}}
\caption{The doping dependence of the particle number fluctuation
$\Delta N_{e}$, the phase fluctuation $\Delta\Theta$, and their
product using (a) the $d$-wave BCS state and (b) the $d$-RVB state
with Jastrow factors in the grand-canonical ensemble. The $d$-wave
BCS state is obtained by solving the BCS Hamiltonian with bare
parameters $(t',t'',V)/t=(-0.2,0.1,6)$ self-consistently. $V$ is the
pairing interaction strength. The blue dashed line denotes 1. (c)
The doping dependence of $\Delta N_{e}$ (squares), $\Delta\Theta$
(circles), and $\Delta N_{e}\Delta\Theta$ (triangles) using the
$d$-RVB state with (filled) and without (empty) Jastrow factors. (d)
The doping dependence of $\frac{1}{\Delta\Theta^{2}N}$ with (filled
diamonds) and without (empty diamonds) Jastrow factors. Black (Red)
color indicates $N=144 \ (256)$. All the results are obtained with
parameters $(t',t'',J)/t=(-0.2,0.1,0.3)$ for a $12\times12$ lattice
except specially mentioned results in (d).} \label{fig.6}
\end{figure}

Before looking into the phase fluctuation $\Delta\Theta$ of the
$d$-RVB wave function, we shall start with the $d$-wave BCS state at
first. To obtain useful information about the phase of wave
functions, a "phase" operator is defined by
\begin{eqnarray}
\hat{\Theta}=\frac{\hat{\Delta}-\hat{\Delta}^{\dag}}{2i\Delta_{0}},
\label{ph2}
\end{eqnarray}
where
$\hat{\Delta}^{\dag}\equiv\sum_{k}\varphi_{k}c_{k\uparrow}^{\dag}c_{-k\downarrow}^{\dag}$
and the normalization
$\Delta_{0}=|\langle\hat{\Delta}^{\dag}\rangle|$. Here we choose
$\varphi_{k}=\langle
c_{k\uparrow}^{\dag}c_{-k\downarrow}^{\dag}\rangle$. Hence, any real
wave function will lead to $\langle\hat{\Theta}\rangle=0$. With a
little algebra, it is straight forward to write down the particle
number fluctuation $\Delta N_{e}$ and the phase fluctuation
$\Delta\Theta$ of the BCS state as
\begin{eqnarray}
\Delta N_{e}&=&2\sqrt{\sum_{k}|u_{k}|^{2}|v_{k}|^{2}},\nonumber\\
\Delta\Theta&=&\frac{\sqrt{\sum_{k}\varphi_{k}^{2}}}{2\sum_{k}\varphi_{k}|u_{k}||v_{k}|}.
\label{ph3}
\end{eqnarray}
Then the uncertainty principle for particle number and phase can be
easily derived by Cauchy-Schwarz inequality:
\begin{eqnarray}
\Delta
N_{e}\Delta\Theta=\sqrt{\frac{\sum_{k}\varphi_{k}^{2}\cdot\sum_{k}|u_{k}|^{2}|v_{k}|^{2}}{\left(\sum_{k}\varphi_{k}|u_{k}||v_{k}|\right)^{2}}}\geq1.
\label{ph4}
\end{eqnarray}
It can be proved that $\Delta N_{e}\Delta\Theta$ in terms of the
definition of Eq.(\ref{ph2}) is exactly equal to 1 in BCS theory.
The doping dependence of the particle number fluctuation and the
phase fluctuation in the $d$-wave BCS case are shown in
Fig.\ref{fig.6}(a).

Although it is impossible to evaluate the phase fluctuation for the
wave function of Eq.(\ref{e:equ1}) with a fixed number of particles,
it is straight forward for the $d$-RVB wave function by calculating
the following quantity
\begin{eqnarray}
\Delta\Theta&\equiv&\sqrt{\left\langle\left(\hat{\Theta}-\left\langle\hat{\Theta}\right\rangle\right)^{2}\right\rangle}\nonumber\\
&=&\sqrt{\frac{\langle\Psi_{d-RVB}^{g}|\hat{\Delta}^{\dag}\hat{\Delta}-\hat{\Delta}^{\dag}\hat{\Delta}^{\dag}|\Psi_{d-RVB}^{g}\rangle}{2\Delta_{0}^{2}}}.\label{ph5}
\end{eqnarray}
In addition to $\Delta_{0}$, there are two quantities to be
calculated by means of the VMC approach: one is the pairing
correlation operator $\hat{\Delta}^{\dag}\hat{\Delta}$ and the other
$\hat{\Delta}^{\dag}\hat{\Delta}^{\dag}$. Using the representation
(df), we can calculate both quantities directly.

In Fig.\ref{fig.6}(b), we show the doping dependence of the particle
number fluctuation and the phase fluctuation using the $d$-RVB wave
function. As mentioned before, the Gutzwiller projection operator
$\hat{P}_{G}$ suppresses the particle number fluctuation as shown in
Fig.\ref{fig.6}(b), especially for the underdoped region. On the
other hand, the phase fluctuation is greatly enhanced for
$0<\delta<0.15$, and a much weaker dependence for doping greater
than $0.15$. Although the fluctuation behavior is approaching BCS
results in the overdoped region, it is still much stronger. This
huge enhancement of $\Delta\Theta$ is clearly due to the strong
correlation effect of the  Gutzwiller projection operator. This
result suggests that there may exist a strong phase-fluctuating
state which is again consistent with experiments in the underdoped
cuprate compounds \cite{YWangPRB06,LeeSci09} and a theoretical
analysis \cite{TesanovicNP08}. Another interesting result is the
large enhancement of $\Delta N_{e}\Delta\Theta$. Instead of having
the value of $1$ as the BCS state, it seems to approach infinity at
very low doping. This is mainly due to the strong increase of phase
fluctuation as doping decreases.

Fig.\ref{fig.6}(c) shows the doping dependence of $\Delta N_{e}$
(squares), $\Delta\Theta$ (circles), and $\Delta N_{e}\Delta\Theta$
(triangles) using the $d$-RVB state with (filled) and without
(empty) Jastrow factors. As discussed before, the Jastrow factor
suppresses the number fluctuation, hence it increases the phase
fluctuation. But the substantial increase of the phase fluctuation
at low doping region is quite surprising since the energy difference
between these two states is quite small as shown in the inset of
Fig.\ref{fig.2}(b). Their product $\Delta N_{e}\Delta\Theta$ for the
case with Jastrow factors also deviates farther from 1 near the
underdoped regime. It indicates that the system in the underdoped
region may have many low energy states with similar energy but
different properties.

In Fig.\ref{fig.6}(d), we find that the doping dependence of
$\frac{1}{\Delta\Theta^{2}N}$ exhibits approximately a linear
relation, shown by the empty diamonds, within $0.03<\delta<0.25$ for
the $d$-RVB state without including the Jastrow factor. Due to the
finite size, it is difficult to get reliable results at extremely
low density but the results of two different cluster sizes are
consistent. This result is in sharp contrast with BCS theory which
has $\frac{1}{\Delta\Theta^{2}N}$ proportional to the pairing gap
instead of the doping density. After the Jastrow factor is included,
as shown by the filled diamonds in Fig.\ref{fig.6}(d), phase
fluctuation is enhanced. However, the linear dependence of doping
density at low density still remains but with a smaller slope.

Empirically the superfluid density of the hole-doped cuprates is
small and proportional to the doping density \cite{UemuraSSC03}. For
a SC state the phase stiffness is proportional the superfluid
density. Although we do not have a direct proof of the relation
between phase stiffness and $\frac{1}{\Delta\Theta^{2}N}$, it is
quite interesting that they both are proportional to the doping
density. In addition, the slope for the wave function with the
Jastrow factor is smaller than the one without the Jastrow factor,
which may indicate that the SC ground state of the $t-J$ model will
have a very small superfluid density just as the hole-doped
cuprates.

\section{Conclusions}

To summarize, we have studied the properties of the ground state and
Bogoliubov quasi-particle states in the extended $t-J$ model based
on Gutzwiller-projected BCS wave function or $d$-RVB wave function
in the grand-canonical ensemble. First of all, by using the phase
space argument used in GA \cite{EdeggerPRB05}, we have numerically
demonstrated how to construct the correct $d$-RVB wave function with
non-conserving particle numbers. A fugacity factor $g$ in front of
$u_{k}$ in the $d$-RVB state is able to efficiently govern the
distribution of empty sites. Our numerical calculations have shown
the excellent agreement obtained for both the optimized grand
thermodynamic potential and tunneling spectra $G(V)$ between the
grand-canonical and the canonical ensemble. It confirms the
necessity and importance of the fugacity factor $g$ in the $d$-RVB
state in the grand-canonical ensemble as emphasized by Anderson
\cite{AndersonLTP06}. The enhanced tunneling asymmetry at low
voltage in the underdoped region is again numerically demonstrated.

In addition, as increasing doping, chemical potential $\mu_{g}$
calculated in the grand-canonical scheme monotonically declines with
the slope which is in good agreement with the experimental results
\cite{HashimotoPRB08}. Almost zero charge susceptibility near
half-filling indicates the incompressible feature due to the Mott
gap. Both the SC order parameter and the particle number fluctuation
have the dome-like doping dependence similar to the SC dome in
high-$T_{c}$ phase diagrams. The doping dependence of the SC order
parameter is similar to that of the particle number fluctuation for
large doping density as the BCS theory. But for low doping density,
there is a significant difference due to the Gutzwiller projection
operator. Furthermore, now we are able to directly calculate the
phase fluctuation in the grand-canonical ensemble. The
Gutzwiller-projected wave function shows not only the smaller
particle number fluctuation but the much enhanced phase fluctuation
than the wave function without Gutzwiller projection in the
underdoped region. We also found $\Delta N_{e}\Delta\Theta$ much
greater than $1$.

In this paper, we only used a uniform fugacity factor in the $d$-RVB
states. Without including a Jastrow factor to obtain a lower energy,
this factor is close to the derived results of GA. However,
including the Jastrow factor produces a much larger fugacity factor
and a significantly enhanced phase fluctuation. This indicates that
the fugacity factor may be quite important in the underdoped region.
In addition, the fugacity factor could have a spatial dependence or
a momentum dependence as noticed by Anderson \cite{AndersonIJMPB11}.
The spatial dependence could produce the stripes as we demonstrated
in Refs.~\onlinecite{CPChouPRB08} and ~\onlinecite{CPC2011} that the
Gutzwiller projection operator introduces the coupling between
charge density, spin density and pair fields. The effect of the
momentum dependence will be left for future study.

\begin{acknowledgments}
We acknowledge stimulating discussions with N. Fukushima, X. M.
Huang, T. Li, Z. Y. Weng, and W. C. Lee. This work was supported by
the National Science Council in Taiwan with Grant No.
98-2112-M-001-017-MY3. The calculations are performed in the
National Center for High-performance Computing and the PC Cluster
III of Academia Sinica Computing Center in Taiwan. F.Y. is grateful
for the NSCF under grant NO.10704008.
\end{acknowledgments}

$^*$yangfan\_blg@bit.edu.cn

\end{document}